\RequirePackage{lineno}

\documentclass[aps,prd,twocolumn,showpacs,superscriptaddress,groupedaddress]{revtex4}

\usepackage{graphicx}% Include figure files
\usepackage{dcolumn}% Align table columns on decimal point
\usepackage{bm}% bold math
\usepackage{verbatim}% Enable \begin{comment} and \end{comment} to comment out text blocks
\begin{document}

% remove the space for publication
%\vspace*{1.5cm}
\hspace{3.2in} \mbox{FERMILAB-PUB-20-568-E;  CERN-EP-2020-236}

%\title{Observation of odderon exchange from combination of a comparison of $pp$ and $p \bar{p}$ elastic scattering with previous $pp$ forward scattering measurement }
\title{%Observation of 
%Odderon exchange from elastic scattering differences between $p \bar{p}$ and $pp$ data at 1.96 TeV and $pp$ forward scattering measurements}
Comparison of $pp$ and $p \bar{p}$ differential elastic cross sections and observation of the exchange of a colorless $C$-odd gluonic compound }

%\input combined_author_list-2.tex
% remove these 3 lines before journal submittal.
%D0 author list dated 10 September 2020, TOTEM author list dated 14 September 2020
%updated for KRC, Gyongyos, Hungary 11 November 2020
%updated for three authors requested by T. Csorgo
% end removal before journal submittal
%
\affiliation{LAFEX, Centro Brasileiro de Pesquisas F\'{i}sicas, Rio de Janeiro, RJ 22290, Brazil}
\affiliation{Universidade do Estado do Rio de Janeiro, Rio de Janeiro, RJ 20550, Brazil}
\affiliation{Universidade Federal do ABC, Santo Andr\'e, SP 09210, Brazil}
\affiliation{INRNE-BAS, Institute for Nuclear Research and Nuclear Energy, Bulgarian Academy of Sciences, Sofia, Bulgaria}
\affiliation{University of Science and Technology of China, Hefei 230026, People's Republic of China}
\affiliation{Universidad de los Andes, Bogot\'a, 111711, Colombia}
\affiliation{University of West Bohemia, Pilsen, Czech Republic}
\affiliation{Charles University, Faculty of Mathematics and Physics, Center for Particle Physics, 116 36 Prague 1, Czech Republic}
\affiliation{Czech Technical University in Prague, 116 36 Prague 6, Czech Republic}
\affiliation{Institute of Physics, Academy of Sciences of the Czech Republic, 182 21 Prague, Czech Republic}
\affiliation{Universidad San Francisco de Quito, Quito 170157, Ecuador}
\affiliation{Helsinki Institute of Physics, University of Helsinki, Helsinki, Finland}
\affiliation{Department of Physics, University of Helsinki, Helsinki, Finland}
\affiliation{LPC, Universit\'e Blaise Pascal, CNRS/IN2P3, Clermont, F-63178 Aubi\`ere Cedex, France}
\affiliation{LPSC, Universit\'e Joseph Fourier Grenoble 1, CNRS/IN2P3, Institut National Polytechnique de Grenoble, F-38026 Grenoble Cedex, France}
\affiliation{CPPM, Aix-Marseille Universit\'e, CNRS/IN2P3, F-13288 Marseille Cedex 09, France}
\affiliation{LAL, Univ. Paris-Sud, CNRS/IN2P3, Universit\'e Paris-Saclay, F-91898 Orsay Cedex, France}
\affiliation{LPNHE, Universit\'es Paris VI and VII, CNRS/IN2P3, F-75005 Paris, France}
\affiliation{IRFU, CEA, Universit\'e Paris-Saclay, F-91191 Gif-Sur-Yvette, France}
\affiliation{IPHC, Universit\'e de Strasbourg, CNRS/IN2P3, F-67037 Strasbourg, France}
\affiliation{IPNL, Universit\'e Lyon 1, CNRS/IN2P3, F-69622 Villeurbanne Cedex, France and Universit\'e de Lyon, F-69361 Lyon CEDEX 07, France}
\affiliation{III. Physikalisches Institut A, RWTH Aachen University, 52056 Aachen, Germany}
\affiliation{Physikalisches Institut, Universit\"at Freiburg, 79085 Freiburg, Germany}
\affiliation{II. Physikalisches Institut, Georg-August-Universit\"at G\"ottingen, 37073 G\"ottingen, Germany}
\affiliation{Institut f\"ur Physik, Universit\"at Mainz, 55099 Mainz, Germany}
\affiliation{Ludwig-Maximilians-Universit\"at M\"unchen, 80539 M\"unchen, Germany}
\affiliation{Department of Atomic Physics, ELTE University, Budapest, Hungary}
\affiliation{E\"otv\"os University, H - 1117 Budapest, P\'azm\'any P. s. 1/A, Hungary}
\affiliation{Wigner Research Centre for Physics, RMKI, Budapest, Hungary}
\affiliation{SzIU KRC, Gy\"ongy\"os, Hungary}
\affiliation{Panjab University, Chandigarh 160014, India}
\affiliation{Delhi University, Delhi-110 007, India}
\affiliation{Tata Institute of Fundamental Research, Mumbai-400 005, India}
\affiliation{University College Dublin, Dublin 4, Ireland}
\affiliation{INFN Sezione di Bari, Bari, Italy}
\affiliation{Dipartimento Interateneo di Fisica di Bari, Bari, Italy}
\affiliation{Dipartimento di Ingegneria Elettrica e dell'Informazione - Politecnico di Bari, Bari, Italy}
\affiliation{INFN Sezione di Genova, Genova, Italy}
\affiliation{Universit\`{a} degli Studi di Genova, Italy}
\affiliation{INFN Sezione di Pisa, Pisa, Italy}
\affiliation{Universit\`{a} degli Studi di Siena and Gruppo Collegato INFN di Siena, Siena, Italy}
\affiliation{Korea Detector Laboratory, Korea University, Seoul, 02841, Korea}
\affiliation{CINVESTAV, Mexico City 07360, Mexico}
\affiliation{Nikhef, Science Park, 1098 XG Amsterdam, the Netherlands}
\affiliation{Radboud University Nijmegen, 6525 AJ Nijmegen, the Netherlands}
\affiliation{AGH University of Science and Technology, Krakow, Poland}
\affiliation{Joint Institute for Nuclear Research, Dubna 141980, Russia}
\affiliation{Institute for Theoretical and Experimental Physics, Moscow 117259, Russia}
\affiliation{Moscow State University, Moscow 119991, Russia}
\affiliation{Ioffe Physical - Technical Institute of Russian Academy of Sciences, St.~Petersburg, Russian Federation}
\affiliation{Institute for High Energy Physics, Protvino, Moscow region 142281, Russia}
\affiliation{Petersburg Nuclear Physics Institute, St. Petersburg 188300, Russia}
\affiliation{Tomsk State University, Tomsk, Russia}
\affiliation{Instituci\'{o} Catalana de Recerca i Estudis Avan\c{c}ats (ICREA) and Institut de F\'{i}sica d'Altes Energies (IFAE), 08193 Bellaterra (Barcelona), Spain}
\affiliation{Department of Astronomy and Theoretical Physics, Lund University, SE-223 62 Lund, Sweden}
\affiliation{Uppsala University, 751 05 Uppsala, Sweden}
\affiliation{CERN, Geneva, Switzerland}
\affiliation{Istanbul University, Istanbul, Turkey}
\affiliation{Taras Shevchenko National University of Kyiv, Kiev, 01601, Ukraine}
\affiliation{Lancaster University, Lancaster LA1 4YB, United Kingdom}
\affiliation{Imperial College London, London SW7 2AZ, United Kingdom}
\affiliation{The University of Manchester, Manchester M13 9PL, United Kingdom}
\affiliation{University of Arizona, Tucson, Arizona 85721, USA}
\affiliation{University of California Riverside, Riverside, California 92521, USA}
\affiliation{SLAC National Accelerator Laboratory, Stanford CA, USA}
\affiliation{Florida State University, Tallahassee, Florida 32306, USA}
\affiliation{Fermi National Accelerator Laboratory, Batavia, Illinois 60510, USA}
\affiliation{University of Illinois at Chicago, Chicago, Illinois 60607, USA}
\affiliation{Northern Illinois University, DeKalb, Illinois 60115, USA}
\affiliation{Northwestern University, Evanston, Illinois 60208, USA}
\affiliation{Indiana University, Bloomington, Indiana 47405, USA}
\affiliation{Purdue University Calumet, Hammond, Indiana 46323, USA}
\affiliation{University of Notre Dame, Notre Dame, Indiana 46556, USA}
\affiliation{Iowa State University, Ames, Iowa 50011, USA}
\affiliation{University of Kansas, Lawrence, Kansas 66045, USA}
\affiliation{Louisiana Tech University, Ruston, Louisiana 71272, USA}
\affiliation{Northeastern University, Boston, Massachusetts 02115, USA}
\affiliation{University of Michigan, Ann Arbor, Michigan 48109, USA}
\affiliation{Michigan State University, East Lansing, Michigan 48824, USA}
\affiliation{University of Mississippi, University, Mississippi 38677, USA}
\affiliation{University of Nebraska, Lincoln, Nebraska 68588, USA}
\affiliation{Rutgers University, Piscataway, New Jersey 08855, USA}
\affiliation{Princeton University, Princeton, New Jersey 08544, USA}
\affiliation{State University of New York, Buffalo, New York 14260, USA}
\affiliation{University of Rochester, Rochester, New York 14627, USA}
\affiliation{State University of New York, Stony Brook, New York 11794, USA}
\affiliation{Brookhaven National Laboratory, Upton, New York 11973, USA}
\affiliation{Case Western Reserve University, Dept.~of Physics, Cleveland, OH 44106 , USA}
\affiliation{Langston University, Langston, Oklahoma 73050, USA}
\affiliation{University of Oklahoma, Norman, Oklahoma 73019, USA}
\affiliation{Oklahoma State University, Stillwater, Oklahoma 74078, USA}
\affiliation{Oregon State University, Corvallis, Oregon 97331, USA}
\affiliation{Brown University, Providence, Rhode Island 02912, USA}
\affiliation{University of Texas, Arlington, Texas 76019, USA}
\affiliation{Southern Methodist University, Dallas, Texas 75275, USA}
\affiliation{Rice University, Houston, Texas 77005, USA}
\affiliation{University of Virginia, Charlottesville, Virginia 22904, USA}
\affiliation{University of Washington, Seattle, Washington 98195, USA}
\author{V.M.~Abazov$^{\dag}$} \affiliation{Joint Institute for Nuclear Research, Dubna 141980, Russia}
\author{B.~Abbott$^{\dag}$} \affiliation{University of Oklahoma, Norman, Oklahoma 73019, USA}
\author{B.S.~Acharya$^{\dag}$} \affiliation{Tata Institute of Fundamental Research, Mumbai-400 005, India}
\author{M.~Adams$^{\dag}$}$^{\dag}$ \affiliation{University of Illinois at Chicago, Chicago, Illinois 60607, USA}
\author{T.~Adams$^{\dag}$} \affiliation{Florida State University, Tallahassee, Florida 32306, USA}
\author{J.P.~Agnew$^{\dag}$} \affiliation{The University of Manchester, Manchester M13 9PL, United Kingdom}
\author{G.D.~Alexeev$^{\dag}$} \affiliation{Joint Institute for Nuclear Research, Dubna 141980, Russia}
\author{G.~Alkhazov$^{\dag}$} \affiliation{Petersburg Nuclear Physics Institute, St. Petersburg 188300, Russia}
\author{A.~Alton$^{a}$$^{\dag}$} \affiliation{University of Michigan, Ann Arbor, Michigan 48109, USA}
\author{ G.A.~Alves$^{\dag}$} \affiliation{LAFEX, Centro Brasileiro de Pesquisas F\'{i}sicas, Rio de Janeiro, RJ 22290, Brazil}
\author{G.~Antchev$^{\ddag}$} \affiliation{INRNE-BAS, Institute for Nuclear Research and Nuclear Energy, Bulgarian Academy of Sciences, Sofia, Bulgaria} 
\author{A.~Askew$^{\dag}$} \affiliation{Florida State University, Tallahassee, Florida 32306, USA}
\author{P.~Aspell$^{\ddag}$} \affiliation{CERN, Geneva, Switzerland}  
\author{A.C.S.~Assis~Jesus$^{\dag}$} \affiliation{Universidade do Estado do Rio de Janeiro, Rio de Janeiro, RJ 20550, Brazil} 
% ...
$^{\dag}$ \affiliation{Lancaster University, Lancaster LA1 4YB, United Kingdom}
\author{C.~Royon$^{\dag}$$^{\ddag}$} \affiliation{University of Kansas, Lawrence, Kansas 66045, USA}
\author{P.~Rubinov$^{\dag}$} \affiliation{Fermi National Accelerator Laboratory, Batavia, Illinois 60510, USA}
\author{R.~Ruchti$^{\dag}$} \affiliation{University of Notre Dame, Notre Dame, Indiana 46556, USA}
\author{G.~Ruggiero$^{\ddag}$} \affiliation{CERN, Geneva, Switzerland}  
\author{H.~Saarikko$^{\ddag}$} \affiliation{Helsinki Institute of Physics, University of Helsinki, Helsinki, Finland}\affiliation{Department of Physics, University of Helsinki, Helsinki, Finland}   
\author{G.~Sajot$^{\dag}$} \affiliation{LPSC, Universit\'e Joseph Fourier Grenoble 1, CNRS/IN2P3, Institut National Polytechnique de Grenoble, F-38026 Grenoble Cedex, France}
\author{V.D.~Samoylenko$^{\ddag}$} \affiliation{Institute for High Energy Physics, Protvino, Moscow region 142281, Russia}
\author{A.~S\'anchez-Hern\'andez$^{\dag}$} \affiliation{CINVESTAV, Mexico City 07360, Mexico}
\author{M.P.~Sanders$^{\dag}$} \affiliation{Ludwig-Maximilians-Universit\"at M\"unchen, 80539 M\"unchen, Germany}
\author{ A.~Santoro$^{\dag}$} \affiliation{Universidade do Estado do Rio de Janeiro, Rio de Janeiro, RJ 20550, Brazil}
\author{A.S.~Santos$^{h}$$^{\dag}$} \affiliation{LAFEX, Centro Brasileiro de Pesquisas F\'{i}sicas, Rio de Janeiro, RJ 22290, Brazil}
\author{G.~Savage$^{\dag}$} \affiliation{Fermi National Accelerator Laboratory, Batavia, Illinois 60510, USA}
\author{M.~Savitskyi$^{\dag}$} \affiliation{Taras Shevchenko National University of Kyiv, Kiev, 01601, Ukraine}
\author{L.~Sawyer$^{\dag}$} \affiliation{Louisiana Tech University, Ruston, Louisiana 71272, USA}
\author{T.~Scanlon$^{\dag}$} \affiliation{Imperial College London, London SW7 2AZ, United Kingdom}
\author{R.D.~Schamberger$^{\dag}$} \affiliation{State University of New York, Stony Brook, New York 11794, USA}
\author{Y.~Scheglov$^{\S}$$^{\dag}$} \affiliation{Petersburg Nuclear Physics Institute, St. Petersburg 188300, Russia}
\author{H.~Schellman$^{\dag}$} \affiliation{Oregon State University, Corvallis, Oregon 97331, USA} \affiliation{Northwestern University, Evanston, Illinois 60208, USA}
\author{M.~Schott$^{\dag}$} \affiliation{Institut f\"ur Physik, Universit\"at Mainz, 55099 Mainz, Germany}
\author{C.~Schwanenberger$^{c}$$^{\dag}$} \affiliation{The University of Manchester, Manchester M13 9PL, United Kingdom}
\author{R.~Schwienhorst$^{\dag}$} \affiliation{Michigan State University, East Lansing, Michigan 48824, USA}
\author{A.~Scribano$^{\ddag}$} \affiliation{INFN Sezione di Pisa, Pisa, Italy}
\author{J.~Sekaric$^{\dag}$} \affiliation{University of Kansas, Lawrence, Kansas 66045, USA}
\author{H.~Severini$^{\dag}$} \affiliation{University of Oklahoma, Norman, Oklahoma 73019, USA}
\author{E.~Shabalina$^{\dag}$} \affiliation{II. Physikalisches Institut, Georg-August-Universit\"at G\"ottingen, 37073 G\"ottingen, Germany}
\author{V.~Shary$^{\dag}$} \affiliation{IRFU, CEA, Universit\'e Paris-Saclay, F-91191 Gif-Sur-Yvette, France}
\author{S.~Shaw$^{\dag}$} \affiliation{The University of Manchester, Manchester M13 9PL, United Kingdom}
\author{A.A.~Shchukin$^{\dag}$} \affiliation{Institute for High Energy Physics, Protvino, Moscow region 142281, Russia}
\author{O.~Shkola$^{\dag}$} \affiliation{Taras Shevchenko National University of Kyiv, Kiev, 01601, Ukraine}
\author{V.~Simak$^{\S}$$^{\dag}$} \affiliation{Czech Technical University in Prague, 116 36 Prague 6, Czech Republic}
\author{J.~Siroky$^{\ddag}$} \affiliation{University of West Bohemia, Pilsen, Czech Republic} 
\author{P.~Skubic$^{\dag}$} \affiliation{University of Oklahoma, Norman, Oklahoma 73019, USA}
\author{P.~Slattery$^{\dag}$} \affiliation{University of Rochester, Rochester, New York 14627, USA}
\author{J.~Smajek$^{\ddag}$} \affiliation{CERN, Geneva, Switzerland} 
\author{W.~Snoeys$^{\ddag}$} \affiliation{CERN, Geneva, Switzerland} 
\author{G.R.~Snow$^{\S}$$^{\dag}$} \affiliation{University of Nebraska, Lincoln, Nebraska 68588, USA}
\author{J.~Snow$^{\dag}$} \affiliation{Langston University, Langston, Oklahoma 73050, USA}
\author{S.~Snyder$^{\dag}$} \affiliation{Brookhaven National Laboratory, Upton, New York 11973, USA}
\author{S.~S{\"o}ldner-Rembold}$^{\dag}$ \affiliation{The University of Manchester, Manchester M13 9PL, United Kingdom}
\author{L.~Sonnenschein$^{\dag}$} \affiliation{III. Physikalisches Institut A, RWTH Aachen University, 52056 Aachen, Germany}
\author{K.~Soustruznik$^{\dag}$} \affiliation{Charles University, Faculty of Mathematics and Physics, Center for Particle Physics, 116 36 Prague 1, Czech Republic}
\author{J.~Stark$^{\dag}$} \affiliation{LPSC, Universit\'e Joseph Fourier Grenoble 1, CNRS/IN2P3, Institut National Polytechnique de Grenoble, F-38026 Grenoble Cedex, France}
\author{N.~Stefaniuk$^{\dag}$} \affiliation{Taras Shevchenko National University of Kyiv, Kiev, 01601, Ukraine}
\author{R.~Stefanovitch$^{\ddag}$} \affiliation{CERN, Geneva, Switzerland} 
\author{A.~Ster$^{\ddag}$} \affiliation{Wigner Research Centre for Physics, RMKI, Budapest, Hungary}
\author{D.A.~Stoyanova$^{\dag}$} \affiliation{Institute for High Energy Physics, Protvino, Moscow region 142281, Russia}
\author{M.~Strauss$^{\dag}$} \affiliation{University of Oklahoma, Norman, Oklahoma 73019, USA}
\author{L.~Suter$^{\dag}$} \affiliation{The University of Manchester, Manchester M13 9PL, United Kingdom}
\author{P.~Svoisky$^{\dag}$} \affiliation{University of Virginia, Charlottesville, Virginia 22904, USA}
\author{I.~Szanyi$^{\ddag}$} \affiliation{Wigner Research Centre for Physics, RMKI, Budapest, Hungary} \affiliation{E\"otv\"os University, H - 1117 Budapest, P\'azm\'any P. s. 1/A, Hungary}
\author{J.~Sziklai$^{\ddag}$} \affiliation{Wigner Research Centre for Physics, RMKI, Budapest, Hungary} 
\author{C.~Taylor$^{\ddag}$} \affiliation{Case Western Reserve University, Dept.~of Physics, Cleveland, OH 44106 , USA}  
\author{E.~Tcherniaev$^{\ddag}$} \affiliation{Tomsk State University, Tomsk, Russia}
\author{M.~Titov$^{\dag}$} \affiliation{IRFU, CEA, Universit\'e Paris-Saclay, F-91191 Gif-Sur-Yvette, France}
\author{V.V.~Tokmenin$^{\dag}$} \affiliation{Joint Institute for Nuclear Research, Dubna 141980, Russia}
\author{Y.-T.~Tsai$^{\dag}$} \affiliation{University of Rochester, Rochester, New York 14627, USA}
\author{D.~Tsybychev$^{\dag}$} \affiliation{State University of New York, Stony Brook, New York 11794, USA}
\author{B.~Tuchming$^{\dag}$} \affiliation{IRFU, CEA, Universit\'e Paris-Saclay, F-91191 Gif-Sur-Yvette, France}
\author{C.~Tully$^{\dag}$} \affiliation{Princeton University, Princeton, New Jersey 08544, USA}
\author{N.~Turini$^{\ddag}$} \affiliation{Universit\`{a} degli Studi di Siena and Gruppo Collegato INFN di Siena, Siena, Italy}
\author{O.~Urban$^{\ddag}$} \affiliation{University of West Bohemia, Pilsen, Czech Republic} 
\author{L.~Uvarov$^{\dag}$} \affiliation{Petersburg Nuclear Physics Institute, St. Petersburg 188300, Russia}
\author{S.~Uvarov$^{\dag}$} \affiliation{Petersburg Nuclear Physics Institute, St. Petersburg 188300, Russia}
\author{S.~Uzunyan$^{\dag}$} \affiliation{Northern Illinois University, DeKalb, Illinois 60115, USA}
\author{V.~Vacek$^{\ddag}$} \affiliation{Czech Technical University in Prague, 116 36 Prague 6, Czech Republic} 
\author{R.~Van~Kooten$^{\dag}$} \affiliation{Indiana University, Bloomington, Indiana 47405, USA}
\author{W.M.~van~Leeuwen$^{\dag}$} \affiliation{Nikhef, Science Park, 1098 XG Amsterdam, the Netherlands}
\author{N.~Varelas$^{\dag}$} \affiliation{University of Illinois at Chicago, Chicago, Illinois 60607, USA}
\author{E.W.~Varnes$^{\dag}$} \affiliation{University of Arizona, Tucson, Arizona 85721, USA}
\author{I.A.~Vasilyev$^{\dag}$} \affiliation{Institute for High Energy Physics, Protvino, Moscow region 142281, Russia}
\author{O.~Vavroch$^{\ddag}$} \affiliation{University of West Bohemia, Pilsen, Czech Republic} 
\author{A.Y.~Verkheev$^{\dag}$} \affiliation{Joint Institute for Nuclear Research, Dubna 141980, Russia}
\author{L.S.~Vertogradov$^{\dag}$} \affiliation{Joint Institute for Nuclear Research, Dubna 141980, Russia}
\author{M.~Verzocchi$^{\dag}$} \affiliation{Fermi National Accelerator Laboratory, Batavia, Illinois 60510, USA}
\author{M.~Vesterinen$^{\dag}$} \affiliation{The University of Manchester, Manchester M13 9PL, United Kingdom}
\author{D.~Vilanova$^{\dag}$} \affiliation{IRFU, CEA, Universit\'e Paris-Saclay, F-91191 Gif-Sur-Yvette, France}
\author{P.~Vokac$^{\dag}$} \affiliation{Czech Technical University in Prague, 116 36 Prague 6, Czech Republic}
\author{H.D.~Wahl$^{\dag}$} \affiliation{Florida State University, Tallahassee, Florida 32306, USA}
\author{C.~Wang$^{\dag}$} \affiliation{University of Science and Technology of China, Hefei 230026, People's Republic of China}
\author{M.H.L.S.~Wang$^{\dag}$} \affiliation{Fermi National Accelerator Laboratory, Batavia, Illinois 60510, USA}
\author{J.~Warchol$^{\S}$$^{\dag}$} \affiliation{University of Notre Dame, Notre Dame, Indiana 46556, USA}
\author{G.~Watts$^{\dag}$} \affiliation{University of Washington, Seattle, Washington 98195, USA}
\author{M.~Wayne$^{\dag}$} \affiliation{University of Notre Dame, Notre Dame, Indiana 46556, USA}
\author{J.~Weichert$^{\dag}$} \affiliation{Institut f\"ur Physik, Universit\"at Mainz, 55099 Mainz, Germany}
\author{J.~Welti$^{\ddag}$} \affiliation{Helsinki Institute of Physics, University of Helsinki, Helsinki, Finland}\affiliation{Department of Physics, University of Helsinki, Helsinki, Finland} 
\author{L.~Welty-Rieger$^{\dag}$} \affiliation{Northwestern University, Evanston, Illinois 60208, USA}
\author{J.~Williams$^{\ddag}$} \affiliation{University of Kansas, Lawrence, Kansas 66045, USA}
\author{M.R.J.~Williams$^{n}$$^{\dag}$} \affiliation{Indiana University, Bloomington, Indiana 47405, USA}
\author{G.W.~Wilson$^{\dag}$} \affiliation{University of Kansas, Lawrence, Kansas 66045, USA}
\author{M.~Wobisch$^{\dag}$} \affiliation{Louisiana Tech University, Ruston, Louisiana 71272, USA}
\author{D.R.~Wood$^{\dag}$} \affiliation{Northeastern University, Boston, Massachusetts 02115, USA}
\author{T.R.~Wyatt$^{\dag}$} \affiliation{The University of Manchester, Manchester M13 9PL, United Kingdom}
\author{Y.~Xie$^{\dag}$} \affiliation{Fermi National Accelerator Laboratory, Batavia, Illinois 60510, USA}
\author{R.~Yamada$^{\dag}$} \affiliation{Fermi National Accelerator Laboratory, Batavia, Illinois 60510, USA}
\author{S.~Yang$^{\dag}$} \affiliation{University of Science and Technology of China, Hefei 230026, People's Republic of China}
\author{T.~Yasuda$^{\dag}$} \affiliation{Fermi National Accelerator Laboratory, Batavia, Illinois 60510, USA}
\author{Y.A.~Yatsunenko$^{\S}$$^{\dag}$} \affiliation{Joint Institute for Nuclear Research, Dubna 141980, Russia}
\author{W.~Ye$^{\dag}$} \affiliation{State University of New York, Stony Brook, New York 11794, USA}
\author{Z.~Ye$^{\dag}$} \affiliation{Fermi National Accelerator Laboratory, Batavia, Illinois 60510, USA}
\author{H.~Yin$^{\dag}$} \affiliation{Fermi National Accelerator Laboratory, Batavia, Illinois 60510, USA}
\author{K.~Yip$^{\dag}$} \affiliation{Brookhaven National Laboratory, Upton, New York 11973, USA}
\author{S.W.~Youn$^{\dag}$} \affiliation{Fermi National Accelerator Laboratory, Batavia, Illinois 60510, USA}
\author{J.M.~Yu$^{\dag}$} \affiliation{University of Michigan, Ann Arbor, Michigan 48109, USA}
\author{J.~Zennamo$^{\dag}$} \affiliation{State University of New York, Buffalo, New York 14260, USA}
\author{T.G.~Zhao$^{\dag}$} \affiliation{The University of Manchester, Manchester M13 9PL, United Kingdom}
\author{B.~Zhou$^{\dag}$} \affiliation{University of Michigan, Ann Arbor, Michigan 48109, USA}
\author{J.~Zhu$^{\dag}$} \affiliation{University of Michigan, Ann Arbor, Michigan 48109, USA}
\author{J.~Zich$^{\ddag}$} \affiliation{University of West Bohemia, Pilsen, Czech Republic} 
\author{K.~Zielinski$^{\ddag}$} \affiliation{AGH University of Science and Technology, Krakow, Poland} 
\author{M.~Zielinski$^{\dag}$} \affiliation{University of Rochester, Rochester, New York 14627, USA}
\author{D.~Zieminska$^{\dag}$} \affiliation{Indiana University, Bloomington, Indiana 47405, USA}
\author{L.~Zivkovic$^{p}$$^{\dag}$} \affiliation{LPNHE, Universit\'es Paris VI and VII, CNRS/IN2P3, F-75005 Paris, France}
%
% visitor_addresses.tex                       10 September 2020
%  available symbols are:
%  $\ast, \dag, \ddag, \S, \P, $\|$, $\ast\ast$, \dag\dag, \ddag\ddag ,\#
%
\collaboration{D0$^\dag$ and TOTEM$^\ddag$ Collaborations\footnote{with visitors from
%{alton}
$^{a}$Augustana University, Sioux Falls, SD 57197, USA,
%{burdin}
$^{b}$The University of Liverpool, Liverpool L69 3BX, UK,
%{grohsjean,deterre}
$^{c}$Deutshes Elektronen-Synchrotron (DESY), Notkestrasse 85, Germany,
%{de la cruz-burelo}
$^{d}$CONACyT, M-03940 Mexico City, Mexico,
%{partridge}
$^{e}$SLAC, Menlo Park, CA 94025, USA,
%{hesketh}
$^{f}$University College London, London WC1E 6BT, UK,
%{luna-garcia}
$^{g}$Centro de Investigacion en Computacion - IPN, CP 07738 Mexico City, Mexico,
%{santos}
$^{h}$Universidade Estadual Paulista, S\~ao Paulo, SP 01140, Brazil,
%{meyer}
$^{i}$Karlsruher Institut f\"ur Technologie (KIT) - Steinbuch Centre for Computing (SCC),
D-76128 Karlsruhe, Germany,
%{patwa}
$^{j}$Office of Science, U.S. Department of Energy, Washington, D.C. 20585, USA,
%{cooke}
%$^{k}$American Association for the Advancement of Science, Washington, D.C. 20005, USA,
%{borysova}
$^{l}$Kiev Institute for Nuclear Research (KINR), Kyiv 03680, Ukraine,
%{jabeen}
$^{m}$University of Maryland, College Park, MD 20742, USA,
%{williams}
$^{n}$European Orgnaization for Nuclear Research (CERN), CH-1211 Geneva, Switzerland,
%{Jung}
$^{o}$Purdue University, West Lafayette, IN 47907, USA,
%{Zivkovic}
$^{p}$Institute of Physics, Belgrade, Belgrade, Serbia,
and
%{Drutskoy}
$^{q}$P.N. Lebedev Physical Institute of the Russian Academy of Sciences, 119991, Moscow, Russia.
%{montgomery}
%$^{?}$Thomas Jefferson National Accelerator Facility, Newport News, VA 23606, USA,
%{falkowski}
%$^{?}$Laboratoire de Physique Theorique, F-91405 Orsay CEDEX, FR,
%{hooper,kozminski}
%$^{?}$}Visitor from Lewis University, Romeoville, IL 60446, USA.
%{weber}
%$^{?}$Universit{\"a}t Bern, CH-3012 Bern, Switzerland.
%{deceased}
%{peng, lipaev, cihangir}
$^{\S}$Deceased.
}} \noaffiliation
\vskip 0.25cm

\date{\today}
%\date{February 21, 2020}
           
\begin{abstract}

We describe an analysis comparing the $p\bar{p}$ elastic cross section as measured by the D0 Collaboration at a center-of-mass energy of 1.96~TeV to that in $pp$ collisions as measured by the TOTEM Collaboration at 2.76, 7, 8, and 13 TeV using a   model-independent approach. The TOTEM cross sections, extrapolated to a center-of-mass energy of $\sqrt{s} =$ 1.96 TeV, are compared with the D0 measurement in the region of the diffractive minimum and the second maximum of the $pp$ cross section.  
The two data sets disagree at the 3.4$\sigma$ level and thus provide evidence for the $t$-channel exchange of a  colorless, $C$-odd gluonic compound, also known as the odderon. 
We combine these results with a TOTEM analysis of the same $C$-odd exchange based on the total cross section and the ratio of the real to imaginary parts of the forward elastic strong interaction scattering amplitude in $pp$ scattering for which the significance is between 3.4 and 4.6$\sigma$.  
The combined significance is larger than 5$\sigma$ and is interpreted as the first observation of the exchange of a colorless, $C$-odd gluonic compound. 
\end{abstract}


\maketitle

%\setpagewiselinenumbers
%\modulolinenumbers[5]
%\linenumbers

The attempts to understand and describe the mechanisms governing the elastic and total cross sections of hadron scattering have evolved over the past seventy years, starting from Heisenberg's observation~\cite{heisenberg} that total cross sections should rise at high energies like $\log^2 s$ where  $s$ is the center of mass energy squared.  This behavior was formalized as the Froissart-Martin bound showing that on very general grounds~\cite{froissard1,froissard2,froissard3}  the total cross section is bounded by $\sigma_{tot} \sim    log^2 s$  as $s$ becomes asymptotically large.                     

Experimental discoveries in the 1970's showed that the $pp$ and $p \bar{p}$ total cross sections at the Intersecting Storage Rings (ISR) do rise with energy~\cite{rise} and can be parametrized with this functional form, albeit with a much smaller constant term than in the Froissart-Martin bound.   The observed experimental rise of $\sigma_{tot}$ with energy has now been extended to much higher $\sqrt{s}$ at the Tevatron, Large Hadron Collider (LHC), and with cosmic rays~\cite{pdg}.  
  
\begin{figure}
\centering
\includegraphics{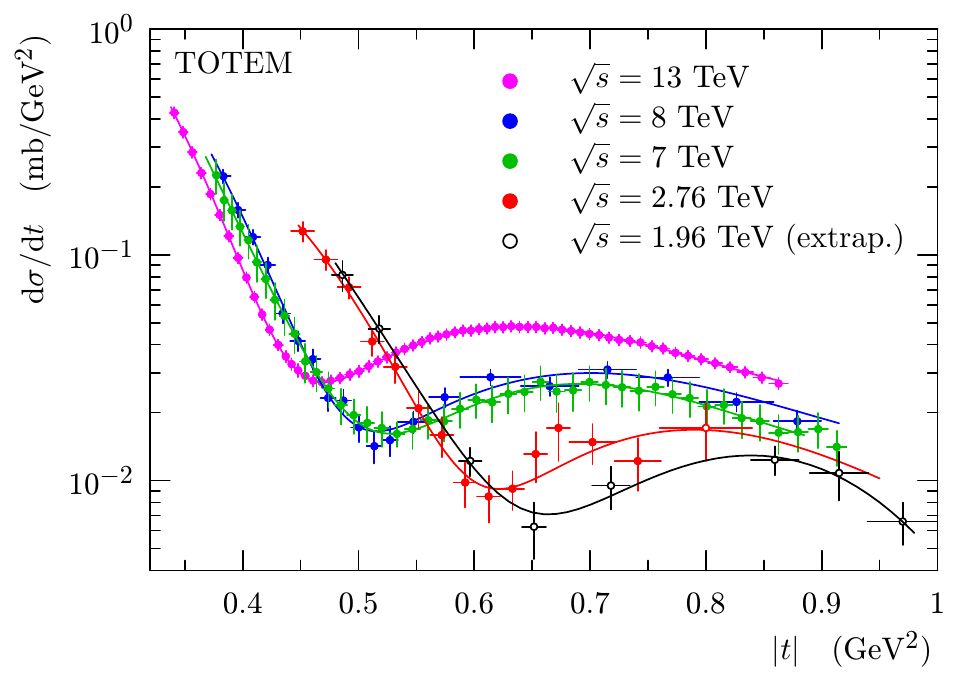}
\caption{ 
The TOTEM measured $pp$ elastic cross sections as  functions of $|t|$ at 2.76, 7, 8, and 13 TeV (full circles), and the extrapolation (discussed in the text) to 1.96 TeV (empty circles). The lines show the double exponential fits to the data points (see text). %The $p \bar{p}$ measurement by the D0 Collaboration at 1.96 TeV is also shown in full triangles.
}
\label{fig:data}
\end{figure}

This behavior was understood in terms of Regge theory in which  $S$-matrix elements for elastic scattering are based on the assumptions of Lorentz invariance, unitarity, and analyticity.  In the high energy Regge limit, the scattering amplitude can be determined by singularities in the complex angular momentum plane.  The simplest examples, Regge poles, lead to terms of the form $\eta  f(t) (s/s_0)^{\alpha(t)}$, where $t$ is the four-momentum transferred squared, $\eta$  the ``signature" with value $\pm1$, and $\alpha(t)$  the ``trajectory" of the particular Regge pole.  Positive signature poles give the same (positive) contribution to both $pp$ and $p\bar{p}$ scattering. Negative signature poles give opposite sign contributions to $pp$ and $p\bar{p}$ scattering.  Using the optical theorem, each such Regge pole contributes a term proportional to $s^{\alpha(0)-1}$ to the total cross section.  The largest contributor at very high energy, called the Pomeron, is the positive signature Regge pole whose $\alpha(0)$ is the largest.  To explain the  rising total cross section, the Pomeron should have $\alpha(0)$ just larger than one.  A $\eta= -1$ Regge exchange with a similarly large $\alpha(0)$, called the odderon~\cite{footnote10,odderon} was recognized as a possibility but was   initially not well motivated theoretically and no clear evidence for it was found~\cite{ISR1,ISR2,footnote11}.

With the advent of Quantum Chromodynamics (QCD) as the theory of the strong interaction, the theoretical underpinnings evolved.    
The exchange of a family of colorless $C$-even states, beginning with a $t$-channel exchange of two gluons, was demonstrated  to play the role of the Pomeron~\cite{pom1,pom2,pom3,nussinov1}  with a predominantly imaginary amplitude near $|t|=$0. QCD also firmly predicted the corresponding predominantly real odderon exchange of a family of colorless $C$-odd states, beginning with a $t$-channel exchange of three gluons, and $\alpha(0)$ near one~\cite{odderon2,nussinov2,odd1,odd2,odd3,bouquet,leader,levin1,levin2,levin3}.  
%At weak coupling, the exchange of compound states of colorless two (or even numbers of) gluons was demonstrated to play the role of the Pomeron~\cite{pom1,pom2,pom3,nussinov1}  with a predominantly imaginary amplitude near $|t|=0$.   QCD also firmly predicted the corresponding predominantly real odderon exchange of compound states of three (or odd numbers of) gluons and $\alpha(0)$ near one~\cite{odderon, odderon2, nussinov1, nussinov2, leader,odd1,odd2,odd3,levin1,levin2,levin3}.    %Several possible variants of Odderons were identified.     
However, the odderon remained elusive experimentally due to the dominating contribution by the Pomeron to total cross sections and small angle elastic scattering.  The effect of the odderon should be felt most strongly when the dominant Pomeron amplitude becomes small compared to the odderon (e.g. near the so-called diffractive minimum in the elastic cross section) leading to an observable difference between $pp$ and $p\bar{p}$ elastic scattering, or in the ratio of the real to imaginary part of the forward strong interaction scattering amplitude.  A recent analysis by the TOTEM collaboration of this ratio and of the total cross section in 13 TeV $pp$ scattering provided strong evidence that the odderon amplitude was needed~\cite{rho}.

This paper presents a model-independent comparison of the $pp$ elastic cross section extrapolated from the measurements  at the LHC to the $p\bar{p}$ cross section measured at the Tevatron.   A difference in these cross sections in the multi-TeV range would constitute a direct demonstration for the existence of the odderon.

\begin{figure}
\centering
\includegraphics[width=0.5\textwidth]{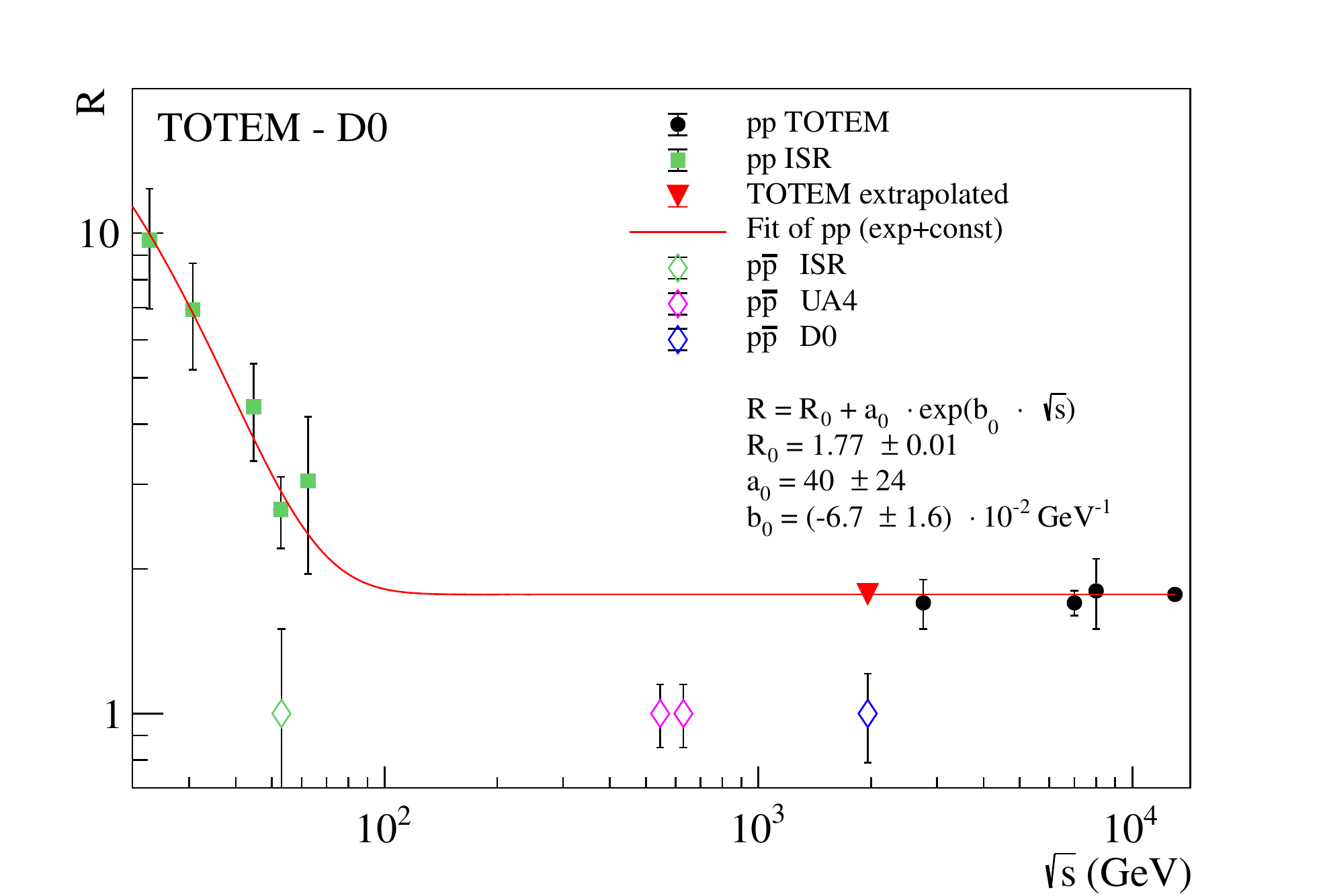}
\caption{
The ratio, $R$, of the cross sections at the bump and dip as a function of $\sqrt{s}$ for $pp$ and $p \bar{p}$. The $pp$ data are fitted to the function noted in the legend.
%The ratio, $R$,  of the cross sections at the bump and dip as a function of $ \sqrt s$.   The points for $ \sqrt{s} <$ 100 (????) GeV are from the ISR measurements.  The $pp$ data are fitted to the function   
%noted in the legend.  The ratio for the D0 and ??? $p \bar{p}$ results are shown at ISR energies and $\sqrt{s} =$ 1.96 TeV.
}
\label{ratioR}
\end{figure}

%\begin{figure}
%\centering
%\includegraphics{referencepoints.pdf}
%\caption{Schematic definition of the characteristic points in the TOTEM differential cross section data. $A$ represents the vertical distance between bump and dip.}
%\label{referencepoints}
%\end{figure}

The D0 Collaboration~\cite{FPD} measured the  $p\bar{p}$ elastic differential cross section at $\sqrt{s}=$1.96 TeV~\cite{d0cross}.  %Scattered particles were detected within the range $0.26 \leq |t| \leq 1.20$ GeV/c$^2$ in scintillating fiber arrays at 23 -- 31 m from the intersection point (IP) after bending in Tevatron magnets.  The systematic uncertainties include effects of beam divergence, detector locations and efficiencies, and the method used for unfolding the measured cross sections.   
The TOTEM Collaboration~\cite{TOTEM} at the CERN LHC measured the differential elastic $pp$ cross sections at $\sqrt{s}$ = 2.76~\cite{totem276}, 7~\cite{totem7}, 8~\cite{totem8} and 13~\cite{totem13} TeV.  %The experiment~\cite{TOTEM} utilizes sets of silicon strip detectors at 213 -- 220 m from the IP to detect elastically and diffractively scattered protons at very small angles after deflections in LHC magnets. 
Figure~\ref{fig:data} shows the TOTEM differential cross sections  used in this study as  functions of $|t|$. %~\cite{footnote3}.   
All $pp$ cross sections show a common pattern of a diffractive minimum (``dip'') followed by a  secondary maximum (``bump'') in $ d\sigma/dt$. 
Figure~\ref{ratioR} shows the ratio $R$ of the differential cross sections measured at the bump and dip locations 
as a function of $\sqrt{s}$ for ISR~\cite{nagy,ISR1}, S$p\bar{p}$S~\cite{SPS1,SPS2}, Tevatron~\cite{d0cross} and LHC~\cite{totem276,totem7,totem8,totem13} $pp$ and $p \bar{p}$ elastic cross section data. The $pp$ data are fitted using the formula 
$R = R_0 + a_0 \exp (b_0 \sqrt{s})$.
We note that the $R$ of $pp$ decreases as a function of $\sqrt{s}$ in the ISR regime and flattens out at LHC energies. Since there is no discernible dip or bump in the D0 $p \bar{p}$ cross section,  we estimate $R$ by taking the maximum ratio of the measured $d\sigma/dt$ values over the three 
neighboring bins centered on the evolution as function of $\sqrt{s}$ of the bump and dip locations as predicted by the $pp$ measurements. The D0 $R=$1.0 $ \pm$ 0.2 value differs from the $pp$ ratio by more than 3$\sigma$ assuming that the flat $R$ behavior of the $pp$ cross section ratio at the LHC continues down to 2 TeV.    The $R$ values shown in Fig.~\ref{ratioR} for $p \bar{p}$  scattering at the ISR~\cite{ISR1} and the S$p\bar{p}$S~\cite{SPS1,SPS2} are similar to those of the D0 measurement.

\begin{figure*}
\centering
\includegraphics[width=0.99\textwidth]{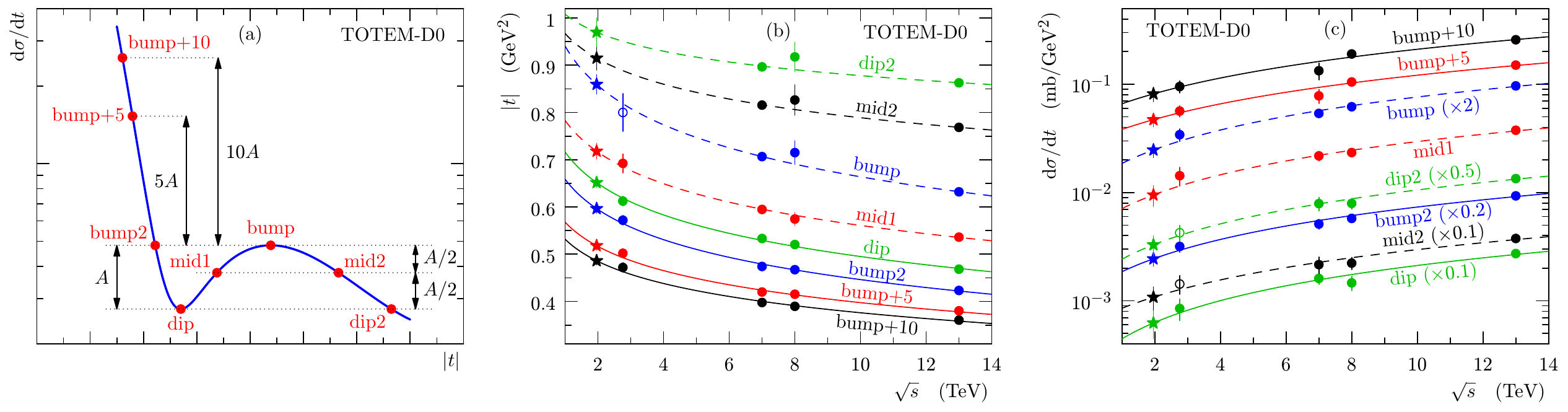}
\caption{$(a)$ Schematic definition of the characteristic points in the TOTEM differential cross section data. $A$ represents the vertical distance between bump and dip. $(b)$ and $(c)$ Characteristic points in $(b)$ $|t|$ and $(c)$ $d\sigma/dt$ from TOTEM measurements at 2.76, 7, 8, and 13 TeV (circles) as a function of $\sqrt{s}$ extrapolated to Tevatron center-of-mass energy (stars).  %using (a) Eq.~\ref{eqn2} and (b) Eq.~\ref{eqn3}.
On $(c)$, a multiplication factor indicated in parenthesis is applied in order to distinguish the different fits. Filled symbols are from measured points; open symbols are from extrapolations or definitions of the characteristic points.}
\label{fig:fits}
\end{figure*}

Motivated by the features of the $pp$ elastic $d\sigma/dt$ measurements, we define a set of eight characteristic points 
as shown in  Fig.~\ref{fig:fits}$(a)$.
For each characteristic point, we identify the values of  $|t|$ and $d \sigma/dt$ at the closest measured points to the characteristic point, thus avoiding
the use of model-dependent fits.
In cases where two adjacent points are of about equal value, the data bins are merged. 
This leads to a distribution of $|t|$ and $d \sigma/dt$ values as a function of $\sqrt{s}$ for all characteristic points as shown in Fig.~\ref{fig:fits}$(b)$ and $(c)$.
The uncertainties correspond to half the bin size in $|t|$ (comparable to the $|t|$-resolution) and to the published uncertainties on the cross sections.

%\begin{figure}
%\end{figure}

The values of $|t|$ and $d \sigma/dt$ as functions of $\sqrt{s}$ for each characteristic point are fitted using the functional forms 
$|t| = a \log (\sqrt{s}) + b  $ and 
$(d\sigma/dt)  = c \sqrt{s} + d $ respectively.
%\begin{eqnarray}
%|t| &=& a \log (\sqrt{s}) + b  \label{eqn2}\\
%(d\sigma/dt)  &=& c \sqrt{s} + d \label{eqn3}.
%\end{eqnarray} 
The parameter values are determined for each characteristic point separately and the same functional form describes the dependence for all characteristic points. The fact that the same forms can be used for all points is not obvious and might be related to general properties of elastic scattering~\cite{nicolescu}.  The $\chi^2$ values for the majority of fits are close to 1 per degree of freedom (dof).  
%Alternative forms that give adequate fits yield extrapolated values that are the same within uncertainties.  
The above forms were chosen for simplicity after it was checked that alternative forms providing adequate fits yielded similar extrapolated values within uncertainties. %  and the above forms were chosen for simplicity among the relatively equal parameterizations. 
%, hence the approach used is essentially model-independent.

The $|t|$  and $d\sigma/dt$ values  for the characteristic points for $pp$ interactions extrapolated to 1.96 TeV are displayed as open black circles  
in Fig.~\ref{fig:data}. The uncertainties on the extrapolated $|t|$ and $d\sigma/dt$ values are computed using a full treatment of the fit uncertainties, taking into account  the fact that the systematic uncertainties of the different characteristic  points  
are not correlated because they correspond to different detectors, data sets and running conditions.

To compare the extrapolated $pp$ elastic cross sections 
with the $p \bar{p}$  measurements, we fit the $pp$ cross section with the function
\begin{eqnarray}
h(t) = a_1 e^{-a_2 |t|^2 - a_3|t|}  
+ a_4 e^{-a_5 |t|^3 - a_6 |t|^2 -a_7 |t|}.
\label{expfit}
\end{eqnarray}
to allow interpolation to the the $t$-values of the D0 measurements in the
range $0.50 \leq |t| \leq 0.96$ GeV$^2$. 
The  fit gives a $\chi^2$ of 0.63 per dof~\cite{footnote4}.    The first exponential in Eq.~(\ref{expfit}) describes the cross section up to the location of the dip, where it falls below the second exponential that describes the asymmetric bump and subsequent falloff. 
This functional form %(with non-zero $a_7$)  
also provides a good fit for the measured $pp$  cross sections at all energies as shown by the fitted functions in Fig.~\ref{fig:data}.  

We evaluate the $pp$ extrapolation uncertainty from Monte Carlo (MC) simulation in which the cross section values of the eight characteristic points are varied within their Gaussian uncertainties and new fits given by Eq.~\ref{expfit} are performed. Fits without a dip and bump position matching the extrapolated values within their uncertainties are rejected, and slope and intercept constraints are used to discard unphysical fits~\cite{footnote1}. 
The MC simulation ensemble provides a Gaussian-distributed $pp$ cross section at each $t$-value. %allowing a 1$\sigma$ uncertainty band to be defined.  
However, the dip and bump matching requirement 
causes the mean of the $pp$ cross section ensemble distribution to deviate from the best-fit cross section obtained above using Eq.~\ref{expfit} with the parameters of Ref.~\cite{footnote4}. 
%However, the dip and bump matching requirement causes the center of the $pp$ cross section band to deviate from the best-fit cross section given by Eq.~\ref{expfit} with the parameters of Ref.~\cite{footnote4}.  
For the $\chi^2$ comparison with the D0 measurements below, we choose %to use the center of the band 
the mean value of the cross section ensemble at each $t$-value with its corresponding Gaussian variance.  
%However, the dip and bump matching requirement causes the center of the band to deviate from the best-fit cross sections. For the $\chi^2$ comparison with the D0 measurements below we choose the central value of the cross section ensemble at each $t$-value with its corresponding Gaussian variance. %to use the center of the band.  %However, the dip and bump matching constraints cause the median of the 
%band  to deviate from the best-fit cross sections.  In accordance with current practice at the LHC and Tevatron in cases
%where the systematic uncertainties are asymmetric, we choose the median
%for the $ \chi^2$ calculation.

\begin{figure}
\centering
\includegraphics{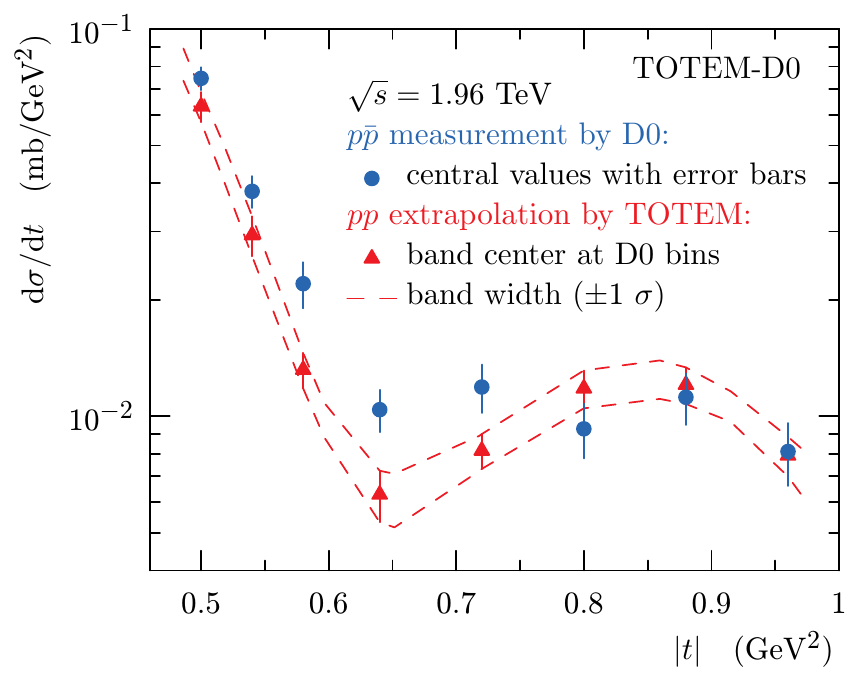}
\caption{Comparison between the D0 $p\bar{p}$ measurement
at 1.96 TeV and the extrapolated TOTEM $pp$ cross section, rescaled to match the OP of
the D0 measurement.  The dashed lines show the
1$\sigma$ uncertainty band on the extrapolated $pp$ cross section.
}
\label{d0andtotem}
\end{figure}

We scale the $pp$ extrapolated cross section so that the optical point (OP), 
$d\sigma/dt(t=0)$, is the same as that for $p \bar{p}$.  The  cross sections at the OP are expected to be   equal if there are only $C$-even exchanges. 
Possible $C$-odd effects~\cite{nicolescu} 
are taken into account below as systematic uncertainties. 
Rescaling the OP for the extrapolated $pp$ cross section would not itself constrain the behavior away from $t=0$.  However, as demonstrated in Refs.~\cite{pomeranchuk,martin} the ratio of the $pp$ and $p \bar{p}$ integrated elastic cross sections becomes one in the limit $\sqrt s \rightarrow \infty$.   The parts of the elastic cross sections in the low $|t|$ Coulomb-nuclear interference region and in the high $|t|$ region above the exponentially falling diffractive cone that do differ for $pp$ and $p \bar{p}$ scattering contribute negligibly to the total elastic cross sections.   Thus, to excellent approximation, the integrated $pp$ and $p \bar{p}$ elastic cross sections in the exponential diffractive region should be the same, implying that the logarithmic slopes should be the same.   As this is the case within uncertainty  for the $pp$ and $p \bar{p}$ cross sections before the OP normalization, we constrain the scaling to preserve the measured logarithmic slopes. We assume that no $t$-dependent scaling beyond the diffractive cone ($|t| \ge$ $0.55$)  is necessary. 

To obtain the OP for $pp$ at 1.96 TeV, we compute the total cross section by extrapolating the TOTEM measurements  at
2.76, 7, 8, and 13 TeV.  
A fit using the functional form%employed since the
%1970's
~\cite{amaldi} for the $s$-dependence of
the total cross section valid only in the range 1 to 13 TeV
\begin{eqnarray}
\sigma_{tot} = b_1 \log^2 (\sqrt{s } / 1 {\rm TeV})  +b_2
\label{equa2}
\end{eqnarray}
gives $\sigma_{tot}^{pp}(1.96$ TeV$)= 82.7 \pm 3.1$ mb~\cite{footnote7}. 
The extrapolated cross section is converted to a differential cross section $d \sigma/dt = 357 \pm 26$ mb/GeV$^2$ at $t=0$ using the optical theorem 
\begin{eqnarray}
\sigma_{tot}^2 = \frac{16 \pi (\hbar c)^2 }{1+\rho^2}  \left( \frac{d \sigma}{dt} (t=0) \right). 
\end{eqnarray}
We assume $\rho = $ 0.145 based on the COMPETE
extrapolation~\cite{compete}. 
%The D0 fit of $d\sigma/dt$ with $0.26<|t|<0.60 GeV$^2$  to a single exponential is extrapolated
The D0 fit of $d \sigma/dt$ for $0.26<|t|<0.6$ GeV$^2$~\cite{d0cross} to a single exponential is extrapolated to $t=0$ to give the OP cross section of $341\pm49$ mb/GeV$^2$.  
Thus the TOTEM OP and extrapolated $d \sigma/dt$ values are rescaled by $0.954 \pm 0.071$  (consistent with the OP uncertainties), where this uncertainty is due to that on the TOTEM extrapolated OP. 
%where the uncertainty is due to that on the TOTEM extrapolated OP as well as possible differences between the $pp$ and $p \bar{p}$ total cross sections (see below).    
We do not claim
that we have performed a measurement of $d\sigma/dt$ at the OP at $t=0$ since this would require additional measurements of
the elastic cross section closer to $t=0$, but we require equal OPs simply  to obtain a common and somewhat arbitrary normalization 
for the two data sets.

The assumption of the equality of the $pp$ and $p \bar{p}$ elastic cross sections at the OP could be modified if an odderon 
exists~\cite{odderon,odderon2}.
A reduction of the significance of a difference between $pp$ and $p \bar{p}$ cross sections would only occur if  the $pp$ total cross section were larger
than the $p\bar{p}$ total cross section at 1.96 TeV.  
This is the case only in maximal odderon  scenarios~\cite{nicolescu}, in which a  1.19 mb difference of the
$pp$ and $p\bar{p}$ total cross sections at 1.96 TeV would correspond to a 2.9\% effect for the OP. This is taken as an
additional systematic uncertainty and added in quadrature to the  quoted OP uncertainty estimated from the TOTEM total cross section fit.  The effect of additional (Reggeon) exchanges~\cite{meson,laszlo,chr}, different methods for  extrapolation to the OP, and potential differences in $\rho$ for $pp$ and $p \bar{p}$ scattering are negligible compared with the uncertainties in the experimental normalization.
The comparison between the extrapolated and rescaled TOTEM $pp$
cross section at 1.96 TeV and the D0 $p\bar{p}$ measurement is shown in Fig.~\ref{d0andtotem} over the interval $0.50 \leq |t| \leq  0.96$ GeV$^2$. 

%We perform a $\chi^2$ test to examine the probability for the D0 and TOTEM differential elastic
%cross sections to agree. The test uses the difference of the scaled cross section in the examined
%$|t|$-range with its fully correlated uncertainty, and the experimental and extrapolated points with
%their covariance matrices.  
%The correlations for the D0 measurements at different $t$-values are small, but the correlations between the eight TOTEM extrapolated data points are large due to the fit using Eq.~\ref{expfit}, particularly for neighboring points. Given the constraints on the OP normalization and logarithmic slopes of the elastic cross sections, the $\chi^2$ test with six degrees of freedom yields the $p$-value of 0.00061, corresponding to a significance of 3.4$\sigma$. 

We perform a $\chi^2$ test to examine the probability for the D0 and TOTEM differential elastic cross sections to agree. The test compares the measured $p \bar{p}$ data points to the rescaled $pp$ data points shown in Fig.~\ref{d0andtotem}, normalized to the integral cross section of the $p \bar{p}$ measurement in the examined $|t|$-range, with their covariance matrices. The fully correlated OP normalization and logarithmic slope of the elastic cross section are added as separate terms to the $\chi^2$ sum. The correlations for the D0 measurements at different $t$-values are small, but the correlations between the eight TOTEM extrapolated data points are large due to the fit using Eq.~\ref{expfit}, particularly for neighboring points. Given the constraints on the normalization and logarithmic slopes, the $\chi^2$ test with six degrees of freedom yields the $p$-value of 0.00061, corresponding to a significance of 3.4$\sigma$.

We make a cross check of this result using an adaptation of the Kolmorogov-Smirnov test in which correlations in uncertainties are taken into account using simulated data sets~\cite{cholesky,choleskyb}.  This cross check,  including the effect of the difference in the integrated cross section in the examined $|t|$-range
via  
the Stouffer method~\cite{stouffer}, gives a $p$-value for the agreement of the $pp$ and $p\bar{p}$ cross sections that is equivalent to the $\chi^2$ test.

We interpret this  difference in the $pp$ and $p \bar{p}$ elastic differential cross sections as evidence that two scattering amplitudes are present and that their relative sign differs for $pp$ and $p\bar{p}$ scattering.  These two processes are even and odd under crossing (or $C$-parity) respectively and are identified as Pomeron and odderon exchanges.  
The dip in the elastic cross section is generally associated with the $t$-value where the Pomeron-dominated imaginary part of the amplitude vanishes. Therefore the odderon, believed to constitute a significant fraction of the real part of the amplitude, is expected to play a large role at the dip. In agreement with predictions~\cite{nicolescu,valery}, the $pp$ cross section exhibits a deeper dip and stays below the $p \bar{p}$ cross section at least until the bump region.

We combine the present analysis result with independent TOTEM odderon evidence based on the measurements of $\rho$ and $\sigma_{tot}$ for $pp$ interaction at different $\sqrt{s}$. These variables are sensitive to differences in $pp$ and $p \bar{p}$ scattering.  The  $\rho$ and $\sigma_{tot}$ results are incompatible with models with only Pomeron exchange and provide independent evidence of odderon exchange effects~\cite{rho}, based on observations in completely different $|t|$ domains and TOTEM data sets. %, as it is also related to differences between $pp$ and $p \bar{p}$ collisions. 

The significances of the different measurements are combined using the Stouffer method~\cite{stouffer}. % in the order of their sensitivity to odderon exchange, starting from the TOTEM $\rho$ measurement at 13 TeV and the present analysis. 
The $\chi^2$ for the total cross section measurements at 2.76, 7, 8 and 13 TeV is computed with respect to the predictions given from models without odderon exchange~\cite{compete,valery} including also model uncertainties when specified. The same is done separately for the TOTEM $\rho$ measurement at 13 TeV~\cite{new1}.  
Unlike the models of Ref.~\cite{compete}, the model of Ref.~\cite{valery} provides the predicted differential cross section without an odderon contribution, so we choose to use %replace the significance of the present analysis in the final combination by the 4.5$\sigma$ significance obtained from a 
the $\chi^2$ comparison of the model cross section at 1.96 TeV with D0 data instead of the D0-TOTEM comparison~\cite{new2}. %at the measured $|t|$-bins in the same $|t|$-range as analyzed by the present analysis~\cite{new2}.

When a partial combination of the TOTEM $\rho$ and total cross section measurements is done, the combined significance ranges between 3.4 and 4.6$\sigma$ for the different models. The full combination leads to total significances ranging from 5.3 to 5.7$\sigma$ for $t$-channel odderon exchange~\cite{new2} for all the models of Refs.~\cite{compete} and \cite{valery}. In particular, for the model favored by COMPETE ($RRP_{nf}L2_u$)~\cite{compete}, the TOTEM $\rho$ measurement at 13 TeV provides a 4.6$\sigma$ significance~\cite{footnote2}, leading to a total significance of 5.7$\sigma$ for $t$-channel odderon exchange when combined with the present result~\cite{new3}.

In conclusion, we have compared the D0 $p\bar{p}$ elastic cross sections at 1.96 TeV and the TOTEM $pp$ cross sections extrapolated to 1.96 TeV from measurements at   2.76, 7, 8, and 13 TeV using a  model independent method~\cite{footnote6}.  The $pp$ and $p\bar{p}$ cross sections differ with a significance of  3.4$\sigma$, and this stand-alone comparison provides evidence that a $t$-channel exchange of a  colorless $C$-odd gluonic compound, i.e. an odderon, is needed to describe elastic scattering at high energies~\cite{nicolescu}. 
When combined with the result of Ref.~\cite{rho}, the significance is in the range 5.3 to 5.7$\sigma$ and thus constitutes the first experimental observation of the odderon.  

The TOTEM collaboration is grateful to the CERN beam optics development team for the design and the successful commissioning of the different special optics and to the LHC machine coordinators for scheduling the dedicated fills. We acknowledge the support from the collaborating institutions and also NSF (USA), the Magnus Ehrnrooth Foundation (Finland), the Waldemar von Frenckell Foundation (Finland), the Academy of Finland, the Finnish Academy of Science and Letters (The Vilho Yrj\"o and Kalle V\"ais\"al\"a Fund), the Circles of Knowledge Club (Hungary), the NKFIH/OTKA grant K 133046 and the EFOP-3.6.1- 16-2016-00001 grants (Hungary). Individuals have received support from Nylands nation vid Helsingfors universitet (Finland), MSMT CR (the Czech Republic), the J\'anos Bolyai Research Scholarship of the Hungarian Academy of Sciences, the New National Excellence Program of the Hungarian Ministry of Human Capacities and the Polish Ministry of Science and Higher Education Grant No. MNiSW DIR/WK/2018/13.
%The TOTEM collaboration acknowledges useful discussions with V.A. Khoze, E. Martynov, B. Nicolescu and M.G. Ryskin.

The D0 Collaboration has prepared this document using the resources of the Fermi National Accelerator Laboratory (Fermilab),
a U.S. Department of Energy, Office of Science, HEP User Facility. Fermilab is managed by Fermi Research Alliance, LLC (FRA),
acting under Contract No. DE-AC02-07CH11359.

The D0 collaboration thanks the staffs at Fermilab and collaborating institutions,
and acknowledges support from the
Department of Energy and National Science Foundation (United States of America);
Alternative Energies and Atomic Energy Commission and
National Center for Scientific Research/National Institute of Nuclear and Particle Physics  (France);
Ministry of Education and Science of the Russian Federation, 
National Research Center ``Kurchatov Institute" of the Russian Federation, and 
Russian Foundation for Basic Research  (Russia);
National Council for the Development of Science and Technology and
Carlos Chagas Filho Foundation for the Support of Research in the State of Rio de Janeiro (Brazil);
Department of Atomic Energy and Department of Science and Technology (India);
Administrative Department of Science, Technology and Innovation (Colombia);
National Council of Science and Technology (Mexico);
National Research Foundation of Korea (Korea);
Foundation for Fundamental Research on Matter (The Netherlands);
Science and Technology Facilities Council and The Royal Society (United Kingdom);
Ministry of Education, Youth and Sports (Czech Republic);
Bundesministerium f\"{u}r Bildung und Forschung (Federal Ministry of Education and Research) and 
Deutsche Forschungsgemeinschaft (German Research Foundation) (Germany);
Science Foundation Ireland (Ireland);
Swedish Research Council (Sweden);
China Academy of Sciences and National Natural Science Foundation of China (China);
and
Ministry of Education and Science of Ukraine (Ukraine).
%
%Both collaborations acknowledge many discussions with theorists such as V.A. Khoze, E. Martynov, B. Nicolescu and M.G. Ryskin.


\begin{thebibliography}{99}
\bibitem{footnote10} We use lower case ``o" to emphasize that there are several theoretical variants of odderon exchanges.
\bibitem{footnote4}  The six parameters to fit the eight data points are
$a_1=9.9$ mb/GeV$^2$, $a_2=16.13$ GeV$^{-4}$, $a_3=2.02$ GeV$^{-2}$, $a_4=2.75$  $\times$ 10$^{-6}$ mb/GeV$^2$, $a_5=28.57$ GeV$^{-6}$, and $a_6=-35.97$ GeV$^{-4}$. The parameter $a_7$ was fixed to zero
as it was not necessary to get a good data description.
\bibitem{footnote1} A few percent of the fits are discarded due to unphysical slopes deviating more than 10$\sigma$ from the one reconstructed from the characteristic points. The number of fits discarded by the intercept constraint is negligible.
\bibitem{footnote7} The fit gives $\chi^2 =0.27$ for 2 dof, with $b_1=4.63\pm0.72$ mb and $b_2=80.64\pm3.36$ mb. We also checked that a polynomial formula or adding an additional $\log$ term in Eq.~\ref{equa2} leads to similar results well within uncertainties.
\bibitem{new1} When the significances computed from the $\rho$ and total cross section measurements at 13 TeV are combined, values determined from independent data sets are used.
\bibitem{footnote2} Ref.~\cite{rho}  reports a 4.7$\sigma$ difference between the favored model of Ref.~\cite{compete} and the TOTEM measurements. This is reduced to 4.6$\sigma$ when the model uncertainties are taken into account.
\bibitem{new3} Here the freedom provided by the Stouffer method to use a subset (e.g. $\rho$ and present analysis) of the significances to compute the final significance is used
\end{thebibliography}
\end{document}